\documentclass[12pt]{article}
\usepackage{epsfig}

\parskip=\medskipamount % space between paragraphs (TeX)
plus.3em minus.5em \abovedisplayshortskip=.5em plus.2em minus.4em
\renewcommand{\thesection}{\arabic{section}}

\catcode`@=11

\@addtoreset{equation}{section} \@addtoreset{equation}{subsection}
\def\theequation{\ifnum\value{section}=0 \arabic{equation}\ignorespaces
\else \ifnum\value{section}=-1 A.\arabic{equation}\ignorespaces
\else \ifnum\value{subsection}=0
\thesection.\arabic{equation}\ignorespaces \else
\thesection.\arabic{subsection}.\arabic{equation}\ignorespaces
                             \fi
                        \fi
                   \fi}

{\catcode`\'=\active \def'{{}^\bgroup\prim@s}}

\catcode`@=12

\newcommand{\bq}{\begin{equation}}
\newcommand{\be}{\begin{equation}}
\newcommand{\fq}{\end{equation}}
\newcommand{\ee}{\end{equation}}
\newcommand{\bqr}{\begin{eqnarray}}
\newcommand{\beqs}{\begin{eqnarray}}
\newcommand{\fqr}{\end{eqnarray}}
\newcommand{\eeqs}{\end{eqnarray}}

\newcommand{\rf}[1]{(\ref{#1})}

\def\bop#1{\setbox0=\hbox{$#1M$}\mkern1.5mu
    \vbox{\hrule height0pt depth.04\ht0
    \hbox{\vrule width.04\ht0 height.9\ht0 \kern.9\ht0
    \vrule width.04\ht0}\hrule height.04\ht0}\mkern1.5mu}
\begin{document}
\thispagestyle{empty}

\begin{flushright}
\begin{tabular}{l}
% TEP- \\
hep-th/0504173 \\
\end{tabular}
\end{flushright}

\vskip .6in
\begin{center}

{\bf  Tree Amplitudes in Scalar Field Theories}

\vskip .6in

{\bf Gordon Chalmers}
\\[5mm]
% {\em address \\
%      address \\
% Los Angeles, CA } \\

{e-mail: gordon@quartz.shango.com}

\vskip .5in minus .2in

{\bf Abstract}

\end{center}

The tree amplitudes in scalar field theories are presented at all $n$.  
The momentum routing of propagators is given at $n$-point in terms of a specified set of 
numbers, and the mass expansion of the massive theories 
is generated.  A group structure on the diagrams is given.  The tree amplitudes 
can be used to find the effective action.

\vfill\break

\noindent{\it Introduction}

The tree amplitudes of gauge theories have been recently under a large
amount of interest, in view of the simplified derivation using the
weak-weak duality of the gauge theory with a twistor formulation.  In
general the tree amplitudes of quantum field theories are required in
order to find the $L\geq 1$ loop amplitudes  (the general form is 
necessary also in the quantum derivative expansion 
\cite{Chalmers1}-\cite{Chalmers10}).  In this work, scalar
field theories are examined in order to generate the coefficients of
the scattering in the mass expansion, and the general form of the classical 
amplitudes.  

The mass expansion of the scalar graphs is direct to obtain.  An obstacle 
is a compact formula that spells out the momentum routing of the individual 
diagrams.  One way of generating this formula is by iterating the amplitudes 
through attaching trees to the ladder diagrams.  The means presented 
gives a mapping between a set of integers that parameterize the diagram to 
another set of numbers that label the propagators.  (The ladders are effectually 
rooted trees and can be useful for other purposes, such as illustrating symmetry.)

The general momentum is sufficient to also specify massless scalar field 
classical scattering.  The massless $\phi^3$ diagrams and their specification 
are required in order to notate all gauge theory and gravity theory tree-level 
amplitudes.  The latter may be obtained in a non-spinor helicity basis, but 
in closed form via the known string-inspired tree-level rules.  The complete 
specification of the spin-$1$ and spin-$2$ classical scattering is forthcoming 
\cite{ChalmersInPrep}.

In $d=4$ the general form of the perturbatively renormalizable scalar
field theory is

\bqr
{\cal L}= {1\over 2} \bigl( \partial\phi\partial\phi + m^2\phi^2\bigr)
+ {\lambda_3\over 3!} \phi^3 + {\lambda_4\over 4!} \phi^4  \ .
\label{relevantth}
\fqr
In general, within a momentum cutoff formalism, there may be further
interactions,

\bqr
{\cal L}_i = \lambda_6 {\phi^6\over \Lambda^2} + \ldots ,
\fqr
which require a further perturbative renormalization of the relevant
terms in the theory \rf{relevantth}.

The counting of the loop parameters for a $\phi^3$ diagram with insertions 
of $v_{2m}^{(2n)}$ operators $\partial^{2n} \phi^{2m}$ is 

\bqr  
3v + \sum 2m v_{2m}^{(2n)} = 2I + E   
\fqr 
\bqr 
L=I-v+1 - \sum v_{2m}^{(2n)} \qquad L\equiv {\rm Loop~No} \ , 
\label{linecount}
\fqr 
with $v$ the number of $\phi^3$ vertices, $I$ the number of internal lines, 
and $E$ the number of external lines.  The case of $\phi^4$ vertices is 
included by setting $m=2$ and $n=0$.  

A derivation of loop amplitudes in either a coupling expansion involving
the parameters $\lambda_i$ or in momenta require the general form of the
classical scattering, i.e. tree-level scattering.  The general form at all 
$n$-point has not appeared in the literature and is presented here.  The 
$\phi^3$ and $\phi^4$ 
scalar field theories pertinent to four dimensions are examined in this work, 
although general theories may be examined as well.  The latter theories are 
of import to higher dimension operators (i.e. 'irrelevant' ones from the definition 
of the renormalization group flow) and to theories in various dimensions.

\vskip .2in 
\noindent {\it $\phi^3$ Theory}

The classical amplitudes considered are placed in a color ordered form; 
the primary ordering is $(1,\ldots,n)$; the non-colored theory's amplitudes 
are derived by summing the sets permutations.    
A general scalar field theory diagram at tree-level is parameterized 
by the set of propagators at the momenta labeling them.  The diagrams 
are labeled by 

\bqr 
D_\sigma = \lambda^{n-2} \prod {1\over t_{\sigma(i,p)} - m^2} \ , 
\label{phi3diagrams} 
\fqr 
with the Lorentz invariants $t_{\sigma(i,p)}$ defined by, 
 
\bqr  
t_{\sigma(i,p)} = (k_{\sigma(i)}+\ldots + k_{\sigma(i+p-1)})^2 \ .  
\label{momentainv}
\fqr 
The sets of permutations $\sigma$ are what are required in order to specify 
the individual diagrams.  The full sets of combinatorics $\sigma(i,p)$ form 
all of the diagrams.  These combinatorics, for a given uniform mass $m$, change 
between $\phi^3$ and $\phi^4$ theory, but are necessary and sufficient to 
label all of the diagrams.   An example 6-point graph has the collection of the 
numbers as: 

\bqr  
\sigma(1,2)=\sigma(3,2)=\sigma(5,2)=\sigma(1,4)=\sigma(3,4)=\sigma(5,4)=1 \ . 
\label{exampleone} 
\fqr 
These numbers parameterize the diagrams with the momenta on the external 
legs following a cyclic ordering.  

An additional vector $\eta(i)$ is required in order to specify the color ordering.  
As only the primary ordering of $1,\ldots,n$ is considered, this vector is not 
relevant.   

The individual terms in the scattering are parameterized by all of the individual 
momenta flows, found by the diagrams in \rf{phi3diagrams}; these terms are to be 
expanded in momenta, using the mass parameter.  Such an expansion is described in 
the 'effective' action by terms such as  

\bqr 
{\cal L}_{\rm eff} = {\lambda^3\over m^2} \phi^5 + {\lambda^4\over m^3} 
 (\partial^2\phi^3) \phi^3 + \ldots  \ . 
\label{exampleeffact}
\fqr 
In general all operators are found by 
expanding the classical scattering (i.e. the tree diagrams) in derivatives, or rather 
in the mass.  The coefficients and the momenta flow are determined from the diagram's 
momentum structure.  The terms in the action \rf{exampleeffact} can be constrained  
in the placement of the derivatives in the 
individual terms of ${\cal L}_{\rm eff}$; as an example, at five-point only the 
$\phi^2 \partial^2 \phi^3$ are allowed due to the momentum flow of the contributing 
diagrams.  This restriction on the momentum structure becomes more apparent when 
the theory is quantized (classically) together with higher dimension operators.  

\begin{figure}
\begin{center}
\epsfxsize=12cm
\epsfysize=12cm
\epsfbox{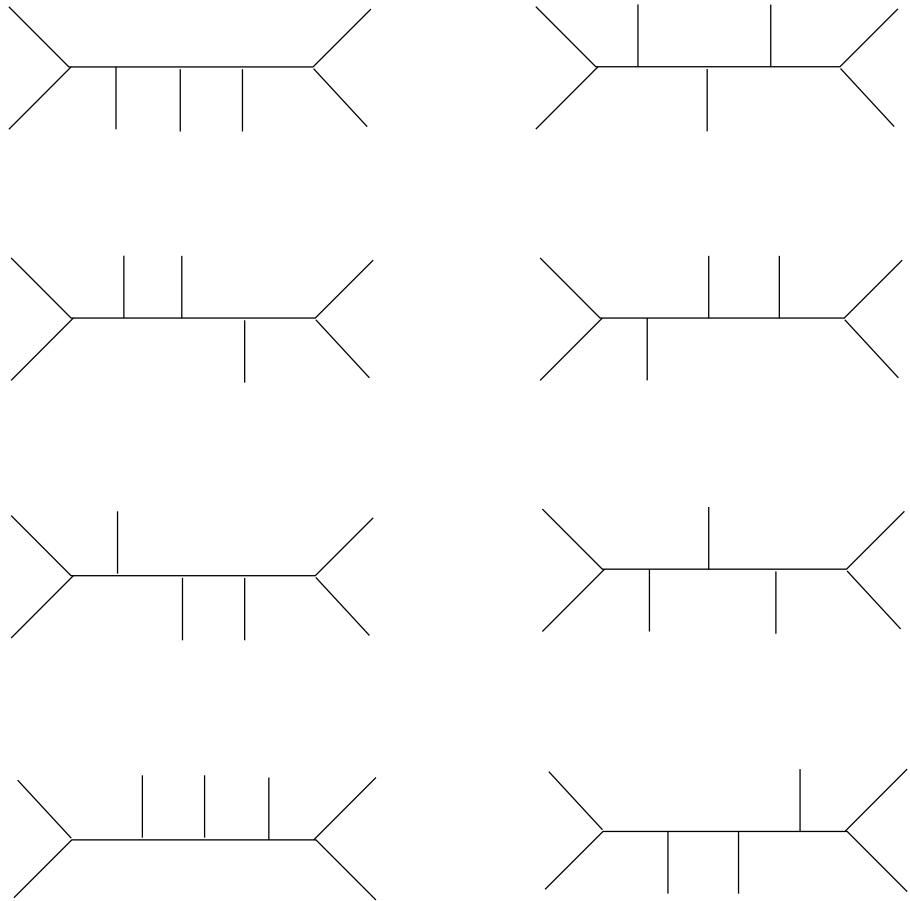}
\end{center}
\caption{The ladder tree diagrams at seven-point.}
\end{figure}

The general term of the classical scattering written in momentum space is, 

\bqr 
D_\sigma = \lambda_3^{n-2} \prod {1\over t_{\sigma_3(i,p)} - m^2} \ .
\label{phi3diagramstwo} 
\fqr
The mass expansion of the diagrams in \rf{phi3diagramstwo}, i.e. with only the 
three-point coupling, is 

\bqr
{\cal L}^n_{\sigma,\tilde\sigma} = C_{\sigma\tilde\sigma}
{\lambda_3^{n-2}\over m^{n-3}}\prod {t_{\sigma(i,p)}^{\tilde\sigma(i,p)}
 \over m^{2\tilde\sigma(i,p)}} \ .
\label{phi3nptclassical}
\fqr 
with the coefficient $C_\sigma$ determined from the tree graphs.  The additional 
set of permutations $\tilde\sigma$ is required in order to specify the expansion 
of the propagators. 

The momenta invariants $t_{\sigma(i,p)}$ at $n$-point are defined in \rf{momentainv}.
In specifying the effective action comprised of all the terms in \rf{phi3nptclassical} 
the numbers and the orderings of $\sigma(i,p)$ (and $\tilde\sigma(i,p)$) must be given.  
In general at $n$-point, the invariants are defined for $i=1$ to
$i=n$ with a non-cyclic ordering of the numbers when $p$ forces the numbers
of $\sigma(j)$ to go beyond $\sigma(n)$ (these numbers pertain to an $n$-point graph).  
For example, at $6$-point the sets of numbers of $\sigma$ are labeled by a collecting the 
indices from $1$ to $n$ in particular orderings such as (e.g. $(1,3,2,6,5,4)$.

The general $\phi^3$ $n$-point diagram, upon color ordering, is constructed from a 
preferred ladder diagram with $n-4$ internal legs.  At each of these legs is attached 
an internal $m<n$ point tree diagram.  The sum of all of these ladder diagrams from 
$m=1$ to $m=n-4$, with the currents attached, generates the complete sum of graphs 
at tree level.  These ladder diagrams at a fixed ordering possess $2^m$ permutations; 
they are illustrated in Figure 2.  The propagator structure of these diagrams can 
be found by iterating the lower-point ladder diagrams.  The preferred basis in 
terms of the ladder trees can be avoided, but some symmetry will be lost in 
in the process.

The internal vertices are labeled so that the outer numbers from a two-particle 
tree are carried into the tree diagram in a manner so that $j>i$ is always chosen 
from the two numbers.  The numbers are carried in from the $n$ most external lines.  
An example diagram with the labeling is illustrated in Figure 3 in the case of 
four-point currents which are to be attached to a ladder tree, and in Figure 4 
for a ladder tree with four internal lines.

The labeling of the vertices is such that in a current the unordered numbers 
are sufficient to reconstruct the current.  For an $m$-point current there are 
$m-1$ vertices and hence $m-1$ numbers contained in $\phi_m(j)$.  These $m-1$ 
numbers are such that the greatest number may occur $m-1$ times, and must occur 
at least once, the next largest number is $m-2$, and so on.  The smallest number 
can not occur in the set contained in $\phi_m(j)$. 

Two example permutation sets are:  

\bqr 
\pmatrix{ 444 \cr 
          443 \cr 
          442 \cr 
          433 \cr 
          432  }
 \label{threeparticle}
\fqr  

\bqr 
\pmatrix{ 5555 \cr 
          5554 \cr 
          5553 \cr 
          5552 \cr 
          5544 \cr 
          5543 \cr 
          5542 \cr 
          5( 3)}
\fqr 
with the $5(3)$ representing the $(3)$-permutation set attached to the $5$ in the 
total count.  There are $5$ and $15$ in the counts. The set of numbers in $\phi(j)$ 
is ordered from largest to least.

The $4+m$ point ladder is labeled as $\kappa=(a_1;a_2;\ldots;a_{m+2})$.  Each of the 
numbers $a_i$ are from the last node of the current; the $a_1$ and $a_{m+2}$ are 
the external 2-point trees.  The vertex numbers are found by the previous 
clockwise $j>i$ labeling; the ladder node numbers $a_j$ are used in this definition.   
(The construction could also be made direct on an ordinary tree graph, without the 
use of currents, but some symmetry would be lost.)

The numbers $\kappa(i)$ and $\phi(j)$ are used to find the propagators in the 
labeled diagram.  The procedure to determine the set of $t_i^{[p]}$, or the 
$\sigma(i,p)$, is as follows.  First, label all momenta as $l_i=k_i$.  Then, 
the invariants are found with the procedure,    

\vskip .2in 
1) $i=\phi(m-1)$, $p=2$, then $l_{a_{m-1}}+l_{a_m}\rightarrow l_{m-1}$ 

2) $i=\phi(m-2)$, $p=2$, then $l_{a_{m-2}}+l_{a_{m-1}}\rightarrow l_{m-2}$ 

... 

$m-1$) $i=1$, $p=m$
\bqr \label{sigmarules} \fqr 
\vskip .2in 

\noindent The labeling of the kinematics, i.e. $t_i^{[p]}$, is direct from the 
definition of the vertices.  

The numbers $\phi_n(i)$ can be arranged into the numbers $(p_i,[p_i])$, in which 
$p_i$ is the repetition of the value of $[p_i]$.  Also, if the number $p_i$ equals 
zero, then $[p_i]$ is not present in $\phi_n$.  These numbers can be used to 
obtain the $t_i^{[q]}$ invariants without intermediate steps with the momenta.  
The branch rules are recognizable as, for a single $t_i^{[q]}$,  

\vskip .2in
0) $l_{\rm initial}=[p_m]-1$

1) 

$r=1$ to $r=p_m$  

${\rm if~} r + \sum_{j=l}^{m-1} p_j = [p_m]-l_{\rm initial}   \quad {\rm then}~i= [p_m] 
  \quad q= [p_m] - l_{\rm initial}+1$ 

beginning conditions has no sum in $p_j$

2) 

${\rm else~} \quad l_{\rm initial}\rightarrow l_{\rm initial}-1$ : 
decrement the line number

$l_{\rm initial}>[p_{l}]$ else $l\rightarrow l-1$ : decrement the $p$ sum 

3) ${\rm goto}~ 1)$ 

\bqr  
\label{branchrules}
\fqr  
The branch rule has to be iterated to obtain all of the poles. 
This rule checks the number of vertices and matches to compare if there is a 
tree on it in a clockwise manner.  If not, then the external line number $l_{initial}$ 
is changed to $l_{initial}$ and the tree is checked again.  The $i$ and $q$ are labels 
to $t_i^{[q]}$.  

The algorithm in \rf{branchrules} has other forms.  There should be a matrix 
transformation between the data in $\phi_n(i)$ to the set of numbers in $t_i^{[p]}$.  
The latter is of dimension $n-3$ (could be twice that if the redundant invariants 
$t_{i+p}^{[n-p]}$ are included in the set) and the former of $n-2$.   

The previous recipe pertains to currents, i.e. amplitudes with one leg off-shell 
and without a line factor.  There are $m-1$ poles in an $m$-point current (does 
not include the off-shell line in $m$, but does include the pole).  In order to 
apply the recipe to an amplitude, the three-point vertex is attached; the counting 
is clear when the attached vertex has two external lines with numbers less than the 
smallest external line number of the current.  There are $n-3$ poles in an $n$-point 
diagram, and these lines are accounted for in the amplitude with this formula; the 
ladder diagrams with their legs can be analyzed with this approach, or simply a 
current with the numbers $\phi_m(i)$.  

An example set of $\sigma(i,p)$ pertains to several seven point diagrams, with 
the indices $i$ and $p$, 

\bqr 
\pmatrix{ 
t_i,2 & t_i,3 & t_i,4 & t_i,5  \cr
16 & 15 & 14 & 13 \cr 
146 & 3 & 6 & 13 \cr 
15 & 47 & 37 & 37 \cr
135 & 5 & 1 & 357 } 
\label{sevenpointkinematics}  \ . 
\fqr 
The vertex labels are, 

\bqr 
\pmatrix{ 75432 \cr 77552 \cr 76662 \cr 77642 }  \ .
\fqr 

\begin{figure}
\begin{center}
\epsfxsize=12cm
\epsfysize=6cm
\epsfbox{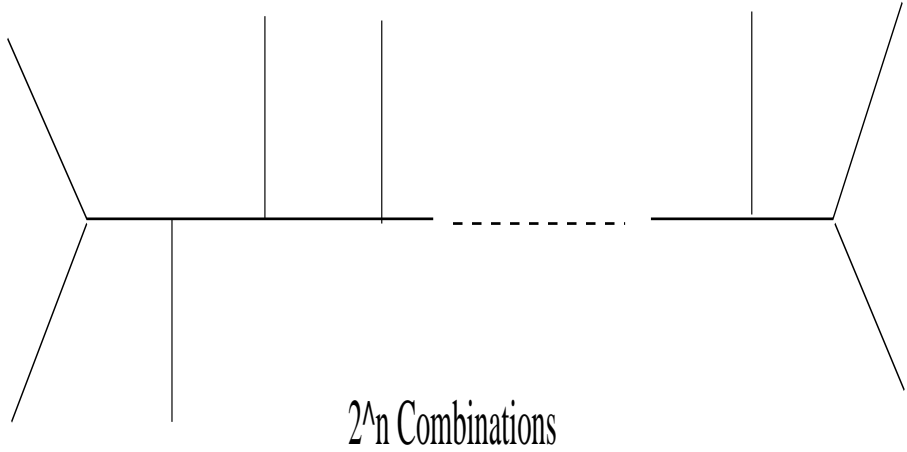}
\end{center}
\caption{The ladder diagrams at $n+4$ point; there are $2^n$ combinations.  Some 
of the 'internal' lines are to be attached to an off-shell current to make the 
various $m>n+4$ point diagrams.}
\end{figure}

\begin{figure}
\begin{center}
\epsfxsize=12cm
\epsfysize=8cm
\epsfbox{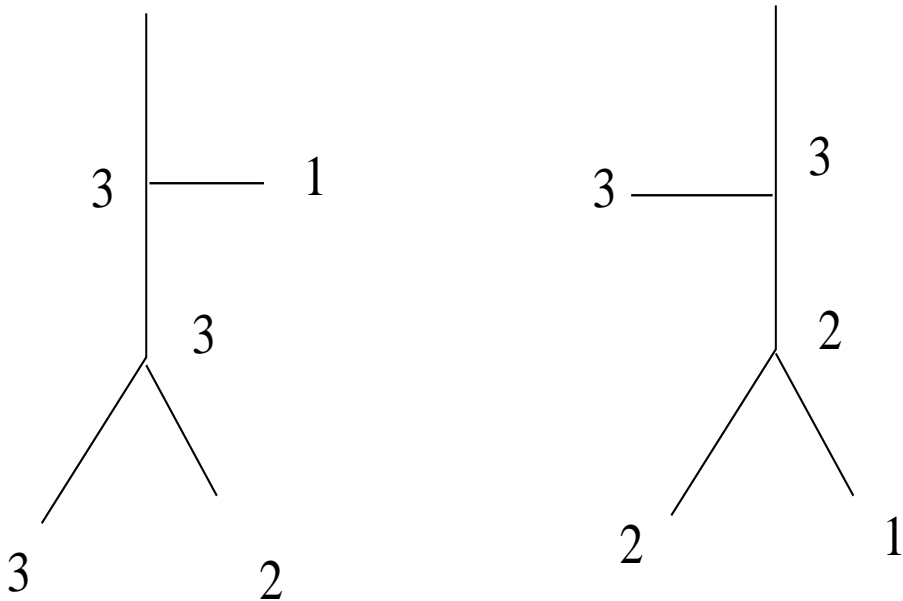}
\end{center}
\caption{The labeling of the 3-particle trees, which are to be sewn on to the 
ladder diagrams.}
\end{figure}

The inclusion of the $\phi^4$ term requires the additional coupling 
$\lambda_4$ (in $d=4$) that alters the coefficients $C_\sigma$.  Higher 
dimension operators such as $\phi^6/\Lambda^2$ for example, require further 
modification of the tree results.  The addition of these higher dimensional 
operators is relevant to model the full quantum field theory and its 
deformations, in general QFTs.

The derivation of the general term in \rf{phi3nptclassical} is obtained
by expanding the general $n$-point graph using the standard propagators
via

\bqr
{1\over  m^2-p^2} = \sum_{q=0}^\infty {1\over m^2} ({p^2\over m^2})^q \ ,
\label{propexp}
\fqr
and collecting the terms in the power series.  Due to the combinatorics 
$\sigma(i,p)$ which parameterize the momenta, the second factor 
$\tilde\sigma(i,p)$ is essentially arbitrary due to the infinite number of 
terms in the expansion \rf{propexp}; the entries are non-vanishing on the 
support of the $\sigma(i,p)$.

\begin{figure}
\begin{center}
\epsfxsize=12cm
\epsfysize=8cm
\epsfbox{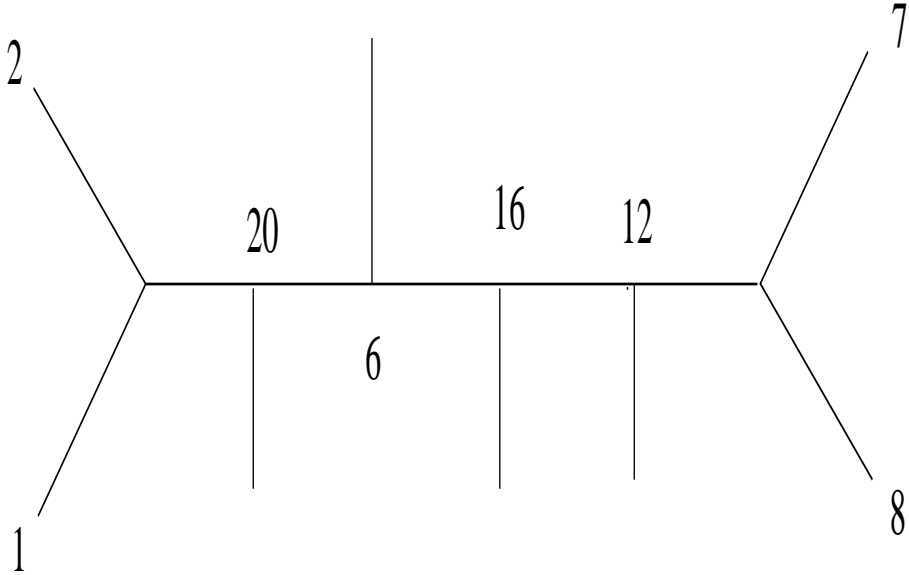}
\end{center}
\caption{The ladder diagram with nodes for four trees.  The labeling is also 
given at these nodes.  The label pertains to the external trees' number.}
\end{figure}

As an example, the zeroeth order term at $n$-point without the derivatives 
is simply found by collecting the number of coupling constants and mass terms
from propagator expansions; the term at $n$-point is,

\bqr
{\cal L}^n_{\sigma,\tilde\sigma} = N_n {\lambda^{n-2}\over m^{2(n-3)}} \ , 
\label{npttwoclassical}
\fqr
with $N_n$ counting the number of graphs at $n$-point order.  
The parameters corresponding to these non-derivative terms are,

\bqr
\tilde\sigma=(0,0,\ldots,0_n) \qquad C_{\sigma,\tilde\sigma}=1 \ ,
\fqr
with $\sigma$ essentially undetermined for the $n$-point amplitude.  The
remaining terms require the expansions of the internal tree propagators,
with momenta following the routings of these lines.

The crux then to finding the general term in \rf{phi3nptclassical} is in the orderings 
of $\sigma(i,p)$.  The coefficients are direct, due to the unity coefficient of 
\rf{propexp} in the expansion of the individual propagators.  The powers in 
the mass expansion are labeled via all integers attached to the propagators, i.e. 
$\tilde\sigma(i,p)$.  The symmetry factors of the diagrams are relevant, in addition to  
the signature of space-time due to the sign in $p^2\pm m^2$.  The set of $\sigma(i,p)$ 
are required to determine all of the terms in the classical effective action. The 
classical 'quantization' could be carried on the basis of the $\sigma(i,p)$, or 
$\phi_n(j)$, as all diagrams are constructable with these vectors.

The general $\sigma_n(i,p)$ can be determined via the collection of numbers in 
$\phi_n(i)$ in \rf{sigmarules} or \rf{branchrules}, and the latter can be labeled 
by the polynomials, 

\bqr  
P(\sigma) = \sum \sigma_n(i,p) y^p x^i \ , 
\label{polynomial}
\fqr 
with the coefficients $(i,p)$ generating the expansion; the coefficients are unity 
as the propagator either exists or not (i.e. zero or one).  The numbers $i$ 
are constrained between $1$ and $n$, and those of $p$ are bounded by $2$ to $n-1$; 
the polynomial is essentially a matrix.  

This polynomial in \rf{polynomial} should satisfy a differential equation due to 
its polynomial nature.  For example, a two-dimensional harmonic oscillator generates 
solutions in terms 
of Hermite polynomials, with a specified degeneracy at level $n_1+n_2$.  The expansion 
of the polynomials in the wavefunction $\psi(x,y) = H_{n_1n_2} e^{-x^2-y^2}$ generate 
sets of coefficients; alternatively an OSp$(n,n)$ wavefunction with the grassmann 
and base expansion could generate the $P(\sigma)$ expansion.  An example differential 
equation at level $n$-point could generate all of the coefficients, and one of the 
form 

\bqr 
\left(\partial_x^2 + \partial_y^2 + V_n(x,y) \right) \psi_n(x,y) = 
   \varepsilon_n \psi_n(x,y) \ ,
\label{wavefunction} 
\fqr  
is desired (or an OSp$(n,n)$ eigenoperator), with degeneracy and form of the wavefunction 
$\psi_n(x,y)$ that of the tree-level field theory diagrams.  The momenta associated with 
the operator in \rf{wavefunction} is obtained with the ordering of the $\sigma$ in a 
cyclic fashion, i.e. the principle color ordered series of $1,2,\ldots, n$. 

The degeneracy of diagrams in field theory for various field theories at $n$-point 
goes as $n!$, and in gauge theory the number of indepedent ones at tree level taking 
into account Ward identities and color orderings follows approximately as $(n-2)!$.  
The specific counting of color ordered $\phi^3$ graphs at $n$-point for low orders 
of $n$ at tree level is,  

\bqr 
\pmatrix{ 
  n=4   &   Q(4)=2  
\cr 
  n=5   &   Q(5)=5 
\cr 
  n=6   &   Q(6)=15
} 
\label{specificphi4}  \ .  
\fqr 
The count of $\phi^4$ graphs at low orders of $n$ is,  

\bqr 
\pmatrix{  
  n=4   &   Q(3)=1  
\cr 
  n=6   &   Q(4)=3  
\cr  
  n=8   &   Q(5)=6 
\cr  
} 
\label{specificphi3}  \ . 
\fqr  
The general count of the tree diagrams follows from the various number theoretic 
combinations of $\phi_n(i)$; that is, the count of sets of $n$ numbers from $n$ 
to $2$ with the maximal occurence of each number being $n-1$ to $1$.  This is 
clearly a factorial.  The count of the $\phi^4$ diagrams is obtained by eliminating 
propagators in $\phi^3$ graphs.

\begin{figure}
\begin{center}
\epsfxsize=12cm
\epsfysize=8cm
\epsfbox{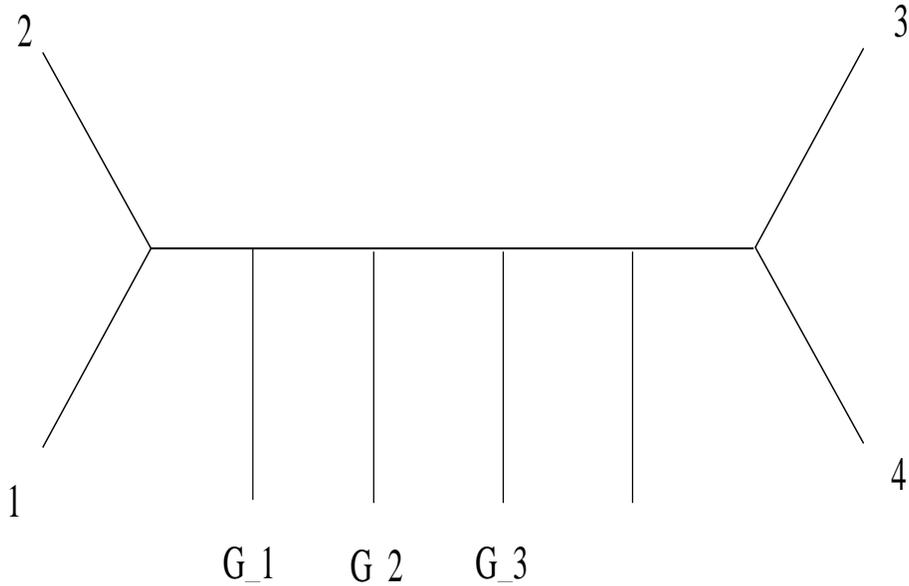}
\end{center}
\caption{The representation of a group structure on the ladder trees.  There are 
$Z_2$ factors at the nodes which orient the current.}
\end{figure}

The algebra associated to tree diagrams can be made clear with the explicit 
parameterization in this section (see \cite{ConnesKreimer}, \cite{Kreimer}  
to find definitions of Hopf structures on rooted trees).  The set of numbers 
$\phi_n(i)$ form a 
basis with an OSp structure.  The $j$th number may repeat at most $j$ times in 
a set of $n-1$ elements; the set of numbers and the combinations is more 
algebraic.  The operations on trees have an action on these vectors $\phi_n(i)$, 
such as eliminating nodes or propagators or exchanging the orders of legs.     
The sets of numbers $\phi_n(i)$ make the Hopf algebra of the rooted trees explicit, 
and the algebra can be obtained without reference to diagrams.  
  
Another structure in the scalar theory is obtained from the currents attached to 
the ladder diagrams, which is not the most general one.  These trees may be 
associated with elements $G_i$ and a $Z_2$ which orients the branch, illustrated 
in Figure 5.  Bracket operations such as $[G_i,G_j]$ can be placed on the trees, 
in accord with the labeling and dimension of the currents' vector space.

Algebras in further classical scalar theories presumably may be obtained by 
pinching the $\phi^3$ diagrams to obtain their graphs, and then using the set 
of numbers $\phi_n(i)$.  The classical quantization can be performed by using 
these numbers directly to obtain the tree diagrams.  The discrete symmetry of 
the set of $\phi_n$, for all $n$, may be understood as an extension of the 
Poincare algebra (which generates the Lagrangian).  It should be useful in 
the quantization.

\vskip .2in 
\noindent {\it $\phi^4$ Theory}

In the case of $\phi^4$ theory the momentum routing of the individual diagrams is 
also modeled by a set of polynomials at $n$-point, 

\bqr  
P(\sigma_n)= \sum \sigma_4(i,p) x^i y^p  \ .  
\fqr 
The determination of the general term in the scattering at tree-level follows in 
almost the same manner, except that the coupling constant has mass dimension in 
$d=4$ and appears with a different factor at $n$-point ($\lambda_4^{n/2-1}$ 
instead of $\lambda_3^{n-2}$).  The color ordered $\phi^4$ diagrams can be obtained  
by pinching the propagators, i.e. removing non-adjacent propagators in certain 
$\phi^3$ trees.  

Several ladder $\phi^4$ diagrams are illustrated in Figure 5.  The pinching 
of $\phi^3$ diagrams is straightforward to obtain.  These ladder diagrams 
are relevant for the same reason as the $\phi^3$ ones are: the potential sewing 
of currents to obtain amplitudes (which is not required), and the manifestation 
of an algebra of the currents and of the theory's classical scattering.  

The diagrams are defined by the $\phi^{(4)}_n(i)$ numbers, which can be found 
by propagator pinching of $\phi^{(3)}_n(i)$.  The pinches occur on every other 
propagator, and this translates to an altering of the set of $\phi^{(3)}_n(i)$.   
 
The expansion in momenta (or in the mass) is accomplished via the 
power expansion of the individual propagators.  The general term in the scattering 
as a function of the kinematic invariants is given by (with $\sigma_4\rightarrow 
\sigma$), 

\bqr
A^n_{\sigma,\tilde\sigma} = C_{\sigma\tilde\sigma}
{\lambda^{n/2-1}\over m^{n/2-2}}\prod {t_{\sigma(i,p)}^{\tilde\sigma(i,p)}
 \over m^{2\tilde\sigma(i,p)}} \ ,
\label{nptclassical}
\fqr

The primary differences between the $\phi^3$ and $\phi^4$ theories are: 1) the 
powers of the coupling and mass differ due to the difference in the number of 
propagators, and 2) the combinatoric factor $\phi^{(4)}_n(i)$ (or $\sigma_4(i,p)$) 
which labels the different momentum routing along the $\phi^4$ diagrams.  

\begin{figure}
\begin{center}
\epsfxsize=12cm
\epsfysize=12cm
\epsfbox{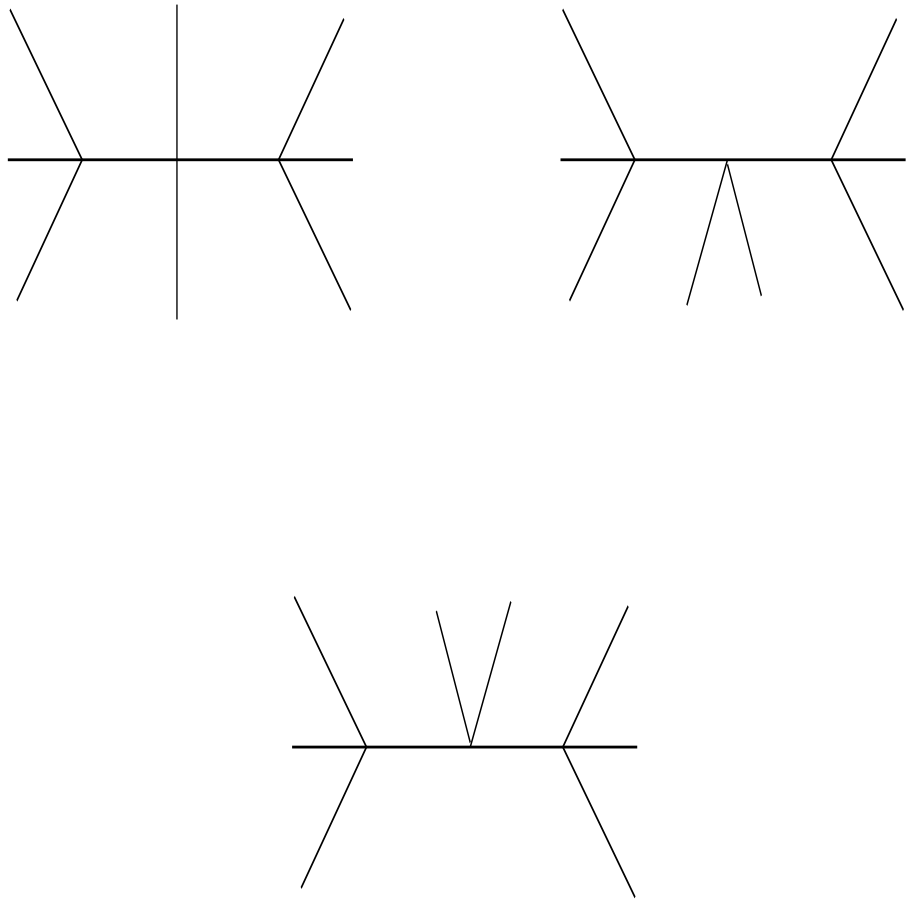}
\end{center}
\caption{The $3$ tree $7$-point ladder diagrams.}
\end{figure}

The $t_i^{[p]}$ invariants which define the propagator define the vector space of 
the polynomials $P(\sigma) = \sum \sigma(i,p) x^i y^j$.  These functions are expected 
to solve, as in $\phi^3$ theory, a differential system, 

\bqr 
\left(\partial_x^2 + \partial_y^2 + V_n^{(3)}(x,y) \right) 
 \psi_n(x,y) = E_n \psi_n(x,y) \ .   
\fqr 
The degeneracy of the solutions should match the number of sets of vertex numbers 
$\sigma_4(i,p)$, which generate the tree amplitudes of the theory.

\vskip .2in 
\noindent {\it $\phi^3$ and $\phi^4$ Theory} 

The combination of $\phi^3$ and $\phi^4$ scalar interactions can be examined in the 
same context as $\phi^4$ theory.  The momentum flow of diagrams in this theory, as 
in the $\phi^3$ theory, can be used to generates the propagator momentum flow in 
gauge theories although this theory is not required for that purpose.  Tuning the 
couplings also interpolates between the two scalar examples. 

The general form of the tree diagrams are more complicated as the number of propagators 
at a given $n$-point varies and is not a function only of the number of external lines 
\rf{linecount}.  For example, at $6$-point there may be one propagator in a pure 
$\phi^4$ graph or three propagators in a $\phi^3$ graph.  The general form of the 
momenta routing is,   

\bqr 
D_\sigma = \lambda_3^{N_3} \lambda_4^{N_4} \prod {1\over t_{\sigma(i,p)} - m^2} \ , 
\label{diagrams} 
\fqr 
with the propagators found via the $\sigma_{3,4}(i,p)$ combinatoric factors.  

The momentum expansion of the diagrams is, 
\bqr
A_{3,4;\sigma,\tilde\sigma}^n = C_{\sigma\tilde\sigma}
{\lambda_3^{N_3} \lambda_4^{N_4}\over m^{N_3-2} }  
  \prod {t_{\sigma(i,p)}^{\tilde\sigma(i,p)} \over m^{2\tilde\sigma(i,p)}} \ .  
\label{nptclassicalphi34}
\fqr
The coefficients $C_{\sigma\tilde\sigma}$ factors, i.e. the momentum tensor factors, 
differ from the individual $\phi_3$, $\phi^4$ theories.  These may be obtained from 
pinching any numbers of propagators in a $\phi^3$ diagram.

The $\sigma(i,p)$ combinatorics is again to be generated by the vector space of 
functions $P_{\sigma_{3,4}}=\sum \sigma_{3,4} x^i y^j$, with degeneracy at level 
$n$ that of the number of $n$-point diagrams.

\vskip .2in 
\noindent {\it Conclusions} 

The set of tree amplitudes in $\phi^3$ and other scalar field theories is given.  
These amplitudes are obtained through sets of numbers $\phi_n$ which describe the 
propagator structure.  The sets of numbers are quite simple; there are $n-1$ of 
them from $2$ to $n$ and the $j$th one cant occur more than $j-1$ times.     
The $\phi^4$ and $\phi^3$ theories, with higher dimensional operators such as 
$\phi^6/\Lambda^2$, are also number theoretic classically; the amplitudes of 
these theories can be found by pinching propagators of the $\phi^3$ theory in 
a systematic fashion, which generates a map to $\phi_n$.  Two dimensional models 
(and other dimensions) can also be examined in this content, in which the 
$\phi^m$ interactions are perturbatively renormalizable.   

Classical scattering of these scalar theories are relevant examples for 
theories with non-vanishing spin.  The routing of the propagators in these 
examples is necessary for the latter theories.  These amplitudes can be 
covariantized in order to find the classical effective action.  

The symmetries of the tree level scattering are obtainable through the vertex 
algebra associated with the diagrams, i.e. $\phi_n(i)$.  The sets of numbers and 
their group aspect can be considered a discrete extension of the Poincare algebra.
These numbers may be used for a direct classical quantization of the scattering, 
as given in the text.  Further sets of numbers, which label the propagators, 
and their symmetries, are necessary for the quantization of higher loops by a 
direct writing down of the loop amplitudes without performing integrals.

\vfill\break


\begin{thebibliography}{99}

\bibitem{ChalmersInPrep} 
G. Chalmers, in preparation.  

\bibitem{Chalmers1}
G. Chalmers, {\it Derivation of Quantum Field Dynamics}, physics/0503062.

\bibitem{Chalmers2}
G. Chalmers, {\it Masses and Interactions of Nucleons Quantum Chromodynamics}, 
physics/0503110.  

\bibitem{Chalmers3}
G. Chalmers, {\it Comment on the Riemann Hypothesis}, physics/0503141.

\bibitem{Chalmers4}
G. Chalmers, {\it $N=4$ Supersymmetric Gauge Theory in the Derivative Expansion}, 
 hep-th/0209088.

\bibitem{Chalmers5}
G. Chalmers, {\it Gauge Theories in the Derivative Expansion}, hep-th/0209086.

\bibitem{Chalmers6}
G. Chalmers, {\it Scalar Field Theory in the Derivative Expansion}, hep-th/0209075.

\bibitem{Chalmers7}
G. Chalmers, {\it M Theory and Automorphic Scattering}, Phys.\ Rev.\ D {\bf 64}:046014 
(2001).

\bibitem{Chalmers8}
G. Chalmers, {\it On the Finiteness of $N=8$ Supergravity}, hep-th/0008162.

\bibitem{Chalmers9}
G. Chalmers and J. Erdmenger, {\it Dual Expansions of $N=4$ super Yang-Mills theory 
via IIB Superstring Theory}, Nucl.\ Phys.\ B {\bf 585}:517 (2000), hep-th/0005192.

\bibitem{Chalmers10}
G. Chalmers, {\it S and U-duality Constraints on IIB S-Matrices}, Nucl.\ Phys.\ B 
{\bf 580}:193 (2000), hep-th/0001190.

\bibitem{ConnesKreimer} 
A. Connes, D. Kreimer, {\it Renormalization in Quantum Field Theory and the 
Riemann-Hilbert Problem}, Comm.\ Math.\ Phys.\ {\bf 210}:249 (2000), hep-th/9912092.    

\bibitem{Kreimer}  
D. Kreimer, {\it Structures in Feynman Graphs: Hopf Algebras and Symmetries}, 
Proc.\ Symp.\ Pure\ Math.\ {\bf 73}:49 (2005), hep-th/022110 ; {\it Combinatorics 
of (Perturbative) Quantum Field Theory}, Phys.\ Rept.\ {\bf 363}:387 (2002), 
hep-th/0010059.   

\end{thebibliography}
\end{document}